\documentclass[5p,sort&compress]{elsarticle}
\usepackage{amssymb}
\usepackage{amsmath}
\usepackage{graphicx}
\usepackage[draft]{hyperref}
\usepackage[utf8]{inputenc}
\usepackage{color}
\usepackage{enumitem}

\renewcommand{\L}{\mathcal{L}}

\renewcommand{\d}{\partial}

\begin{document}

\title{Conformal symmetry: towards the link between the Fermi and the Planck scales}

\author[epfl]{Mikhail Shaposhnikov}
\ead{Mikhail.Shaposhnikov@epfl.ch}
\author[epfl,inr]{Andrey Shkerin}
\ead{Andrey.Shkerin@epfl.ch}

\address[epfl]{
 Institute of Physics, Laboratory for Particle Physics and Cosmology,
\'{E}cole Polytechnique F\'{e}d\'{e}rale de Lausanne, CH-1015 Lausanne, 
Switzerland}

\address[inr]{Institute for Nuclear Research of the Russian Academy
of Sciences, 60th October Anniversary prospect 7a, 117312, Moscow,
Russia}



\begin{abstract}
If the mass of the Higgs boson is put to zero, the classical Lagrangian of the Standard Model (SM) becomes conformally invariant (CI). Taking into account quantum non-perturbative QCD effects violating CI leads to electroweak symmetry breaking  with the scale $v \sim \Lambda_{\rm QCD}\sim 100$ MeV which is three orders of magnitude less than it is observed experimentally. Depending on the mass of the top quark, the radiative corrections may lead to another minimum of the effective potential for the Higgs field with $v \gtrsim M_P$, where $M_P$ is the Planck mass, at least $16$ orders of magnitude more than it is observed. We explore yet another source of CI breaking associated with gravity. We suggest a non-perturbative mechanism that can reproduce the observed hierarchy between the Fermi and the Planck scales, by constructing an instanton configuration contributing to the vacuum expectation value of the Higgs field. The crucial role in this effect is played by the non-minimal coupling of the Higgs field to the Ricci scalar and by the approximate Weyl invariance of the theory for large values of the Higgs field.

\end{abstract}

\maketitle

\section{Introduction and Setup}
\label{sec:Intro}

The observed value of the Electroweak (EW) scale, represented by the Standard Model (SM) Higgs boson mass $m_H$, poses a challenge the physics beyond the SM must deal with. One aspect of this challenge comes from the fact that radiative corrections to $m_H$, engendered by the presence of degrees of freedom with some heavy mass scale (existing, for instance, in the standard GUTs -- Grand Unified theories), shift it towards that scale. The apparent stability of the Higgs mass against such corrections requires either fine-tuning among the parameters of the theory or a special mechanism of their systematic suppression. The second aspect of the problem concerns with the smallness of the ratio of the EW scale to the GUT scale or to the Planck scale  at which quantum gravity effects are expected to come into play.   Combined together, the two issues are known as the hierarchy problem (for reviews see, e.g. \cite{Giudice:2008bi,Giudice:2013nak}).

The first part of the hierarchy problem, related to the stability of the EW scale against perturbative quantum corrections, does not appear in theories  containing  light particles only\footnote{This requirement does not exclude Grand Unified Theories, which can be constructed without leptoquarks  \cite{Karananas:2017mxm}.}, for detailed arguments and previous references see \cite{Vissani:1997ys,Shaposhnikov:2007nj,Farina:2013mla}. Of course, even if one evades any heavy mass thresholds coming with the physics beyond the SM, the Planck scale enters unavoidably any theory comprising the SM and General Relativity (GR). As the Planck mass $M_P$ does not represent a mass of any particle but rather serves as a parameter measuring the strength of gravitational interaction (for an overview see, e.g.  a relevant chapter in \cite{Giudice:2016yja}), the argument may remain in force in the presence of gravity as well \cite{Shaposhnikov:2007nj,Shaposhnikov:2008xi,Farina:2013mla}.

It is tempting to use the conformal symmetry for solution of the second part of the hierarchy problem  \cite{Bardeen:1995kv}. Indeed,  since at the classical level the SM Lagrangian acquires a conformal invariance (CI) once $m_H$ is put to zero, one can imagine to start from the conformally-invariant classical SM which has no EW symmetry breaking and generate the Higgs mass due to the CI violation. 

One of the possible ways to generate the Higgs mass within the CI setting is associated with quantum conformal anomaly (see, e.g., \cite{Coleman:1970je}). Indeed, the UV regularisation of renormalizable field theories necessarily introduces a parameter with the dimension of mass, which violates CI at the quantum level and thus makes it to be anomalous.\footnote{If the requirement of renormalisability is removed, the theory can be made conformal at the quantum level \cite{Englert:1976ep,Shaposhnikov:2008xi,Gretsch:2013ooa}. We will consider such theories in a separate publication.} As a result, the effective potential for the Higgs field, accounting for higher-order radiative corrections, may develop a minimum displaced from the origin, potentially leading to the vacuum expectation value (vev) $v$ of the Higgs field small compared with $M_P$ or the GUT scale \cite{Coleman:1973jx,Weinberg:1978ym}. 

The Coleman-Weinberg (CW) scenario \cite{Coleman:1973jx} in the SM can indeed be realised \cite{Linde:1975sw,Weinberg:1976pe,Linde:1977mm}, but it leads to the Higgs and the top quark masses $m_H \simeq 7$ GeV and  $m_t \lesssim 80$ GeV being far from those observed experimentally. If we take the physical values of dimensionless Higgs self-coupling $\lambda$ and top quark Yukawa coupling $y_t$, the effective potential of the CI SM has a minimum around the QCD scale $\Lambda_{\rm QCD}\sim 100$ MeV, associated with confinement and quark condensates \cite{Witten:1980ez}, which is too far from the one realised in Nature. If the value of the top quark Yukawa coupling is smaller than some critical value,  $y_t<y_{\rm crit}$,  this minimum is unique. For $y_t>y_{\rm crit}$ yet another minimum is generated by the CW mechanism \cite{Froggatt:1995rt}, with the very large vev $v \gtrsim M_P$, now many orders of magnitude larger that the EW scale. Due to experimental uncertainties it is not known yet whether $y_t$ is larger or smaller than $y_{\rm crit}$ (for an overview see, e.g., \cite{Bezrukov:2014ina}), but in any event the predictions of the CI SM are in sharp contrast with experiment.

In spite of this failure, the no-scale CI theories look very attractive and motivated many authors to search for different extensions of the SM, in which the mechanism may work and be phenomenologically acceptable. We mention just a few. The extended scalar sector was discussed on general grounds in \cite{Gildener:1976ih}, more recent works deal with the SM extended by right-handed neutrinos and a real scalar field \cite{Meissner:2006zh}, by an Abelian $B-L$ gauge field \cite{Iso:2009ss,Boyle:2011fq,Helmboldt:2016mpi}, or by non-abelian gauge groups \cite{Karam:2015jta,Karam:2016rsz}. 

All the considerations of the CI theories up to date were carried out without gravity. There is a clear rationale for this, based on (nearly scale-invariant) perturbation theory \cite{tHooft:1972tcz}: any {\em perturbative} corrections to the effective potential of the Higgs field coming from gravity are suppressed by the Planck mass \cite{tHooft:1974toh}, and in the absence of heavy particles they are numerically small. In the SM, the largest contribution is of the order $y_t^6 h^6/M_P^2$, which is negligible at the weak scale.

Clearly, the general difficulty of working with quantum gravity is that we are not aware of an explicit UV completed theory reducing to (CI) SM and GR at low energy scales. What one can do in this situation is to attribute to the unknown UV physics the properties which are found to be useful in resolving apparently low energy physic's questions. In particular, one can imagine that the Higgs mass arises \textit{purely} due to some quantum gravity effects. Having accepted the CI setting, this could only imply the existence of a non-perturbative mechanism driving the Higgs field vev towards its observed value, some 17 orders of magnitude below $M_P$.

The aim of this paper is to argue that gravity can indeed generate in a non-perturbative way a new mass scale, many orders of magnitudes smaller than the Planck scale. Non-perturbative phenomena can manifest themselves in various ways. As one example, they can be associated with the strong coupling scale around which the content of the theory is reorganized, and, in particular, the physical degrees of freedom are rearranged.\footnote{For discussion of this possibility in context with the hierarchy problem see, e.g., \cite{Dvali:2016ovn}.} Another possibility is the existence of euclidean classical configurations --- the instantons --- that contribute to correlation functions of the theory and may eventually result in drastic changes of the low energy observables.\footnote{Perhaps, the most instructive example of this phenomenon is a discrete symmetry restoration in quantum mechanics of one dimension \cite{Polyakov:1987ez}.} In this paper  our main concern will be with the second effect. In fact, it is not difficult to come to an idea that instantons or, more precisely, their large actions may somehow be involved in generating the hierarchy of scales. The smallness of the ratio $v/M_P$ can be viewed as a result of an exponentially strong suppression of the Planck scale that generally appears in physical quantities as the only classical scale of the theory. Our intention is to use some simple yet explicit models comprising the Higgs and the gravitational fields as a playground in which the existence conditions for such instantons can be studied.

Of course, lacking a UV complete theory encompassing the SM and GR brings irremovable ambiguities in our analysis. Nevertheless, we argue that this ambiguity does not make quantitative investigation meaningless once it is clearly stated under what assumptions about the high energy behaviour of the theory we work. We find this kind of approach quite appealing, as the line of reasoning can be reverted easily by claiming the observed value of $v$ to be an argument in favour of those properties of the unknown theory, for which the working mechanism of generation of the Higgs mass is found.


Let us specify the framework in which we will work. Following the discussion above, we would like to exclude from consideration possible quantum (perturbative) corrections to the Higgs field vev, coming with the heavy mass thresholds associated with new physics. To this end, we require no degrees of freedom with the mass scales above the EW scale appear in the theory. That is, we demand the only classical dimensional parameter in the theory be the Planck mass. In this case, the Higgs-gravity sector of a theory under investigation, which we will be mainly interested in, is governed by the Horndeski construction or its extensions \cite{Horndeski:1974wa,Langlois:2015cwa}. The vastness of possible models is further restricted by the requirement to reproduce the SM Higgs sector and GR at low energies and by the assumption that among higher-dimensional operators activating at high energies those are present that we find useful for the purpose of generating the hierarchy of scales.\footnote{The structure of theories at high energies can also be subject to constraints, e.g., by the requirement of asymptotic safety \cite{Weinberg:2009wa}. } It should be stressed that our goal here is to find an example of a model in which a mechanism of an exponential suppression of the Planck mass due to instantons exists. Hence, we do not have an intention to perform a survey of all possible models by studying effects from all possible higher-dimensional operators containing the Higgs and gravity fields. Nor do we intend to argue that a toy theory chosen to illustrate the mechanism can indeed be consistently embedded into the UV complete theory of gravity.


We will find that the crucial ingredient of the theory admitting the instantons with the desired properties is the non-minimal coupling of the Higgs field to the Ricci scalar. We will also find that the instantons generating the large hierarchy of scales favour the (approximate) Weyl invariance of the theory for large values of the scalar field.

The paper is organized as follows. In sec. \ref{sec:Model} we introduce a simple model describing the dynamics of the gravitational and the classically massless scalar fields. We analyze euclidean classical configurations arising in this model, and discuss their possible influence on the vev of the scalar field. The results of the analysis motivate us to introduce certain modifications into the model. In sec. \ref{sec:Mechanism} we incorporate these modifications step by step. We find that the contribution of a certain classical configuration (the singular instanton) to the vev of the scalar field can actually make the latter non-zero and, at the same time, many orders of magnitude smaller than the Planck scale. In sec. \ref{sec:SM} we apply our findings to the EW hierarchy problem by identifying the scalar field with the Higgs field, and discuss the inclusion of other SM degrees of freedom. Finally, sec. \ref{sec:DiscConcl} is dedicated to a general discussion of the non-perturbative mechanism. In particular, we discuss to what extent the toy models studied in sec. \ref{sec:Mechanism} can be generalized without spoiling the mechanism.

\section{The simple model}
\label{sec:Model}

In the euclidean signature, the vev of the scalar field $\varphi$ is evaluated as\footnote{In this paper, we adopt the euclidean the path integral. Because of the presence of gravity, it must be treated with caution \cite{Gibbons:1977zz}. We assume that quantum gravity resolves possible issues arising when using this formulation; see also \cite{Gratton:1999ya}.}
\begin{equation}\label{HiggsVEV}
\langle \varphi\rangle\sim\int\mathcal{D}\mathcal{A}\mathcal{D}\varphi\mathcal{D}g_{\mu\nu}\varphi e^{-S_E} \; .
\end{equation}
Here $\mathcal{A}$ denotes the collection of fields of the model under consideration, other than $\varphi$ and $g_{\mu\nu}$, and $S_E$ is the euclidean action of the model. Should the model admit the classical background of the form $\varphi=0$, $g_{\mu\nu}=\delta_{\mu\nu}$, the conventional perturbation theory built upon it can in principle provide $\langle \varphi\rangle$ with a nonzero value. However, as was discussed in sec. \ref{sec:Intro}, identifying $\varphi$ with the Higgs field and the low energy limit of the model at hand with the (CI) SM and GR, one finds that radiative corrections to the Higgs vev around the flat background do not bring it to the observed value.\footnote{Here we do not discuss the presence of the cosmological constant, since the latter is irrelevant for our analysis. We just note that it can be included into consideration without spoiling the (classical) CI of the theory (see \cite{Shaposhnikov:2008xb,Blas:2011ac,Karananas:2016grc} and references therein).} Therefore, we are interested in possible nontrivial classical configurations that would contribute to the vev $\langle \varphi\rangle$. We restrict the attention to the configurations built from the scalar and the metric fields, while other degrees of freedom present in the theory are kept classically at their vacuum values. Furthermore, we limit the analysis to the main exponential contribution to the path integral in eq. (\ref{HiggsVEV}). This allows us to consider the part of the theory comprising $\varphi$ and $g_{\mu\nu}$ only. In sec. \ref{sec:SM} we will comment on implications of our findings to the SM fields other than the Higgs field.

As a warm-up, in this section we study a simple model containing the real scalar field $\varphi$ coupled to dynamical gravity. The purpose is to elucidate important properties of euclidean classical configurations that appear in the theories with a non-minimal coupling of the scalar field to gravity. We take the following Lagrangian\footnote{The euclidean action $S_E=\int d^4x\mathcal{L}_{\varphi,g}$ of the model must be accompanied with the proper boundary term (see, e.g., \cite{Padilla:2012ze}). Since the latter will not be relevant for our purposes, in what follows we will omit it.}
\begin{equation}\label{L0_J}
\dfrac{\mathcal{L}_{\varphi,g}}{\sqrt{g}}=-\dfrac{1}{2}(M_P^2+\xi\varphi^2)R+\dfrac{1}{2}(\partial\varphi)^2+V(\varphi)
\end{equation}
with 
\begin{equation}\label{L0_Pot}
V(\varphi)=\dfrac{\lambda}{4}\varphi^4
\end{equation}
and $\xi>0$. The scalar sector of the model, represented by the last two terms in eq. (\ref{L0_J}), exhibits global conformal invariance. Addition of gravity and the scalar-gravity interaction (the first two terms) breaks this symmetry explicitly. In the limit $\vert\varphi\vert\gg M_P/\sqrt{\xi}$, the global scale invariance (SI) is acquired. Overall, the model (\ref{L0_J}) serves is a good prototype of the Higgs-gravity sector of a theory we are eventually interested in. Of course, in a more realistic setting operators of higher dimensions suppressed by a proper cutoff must be added to the Lagrangian (\ref{L0_J}). We will proceed with the study of particular types of such operators in sec. \ref{sec:Mechanism}. Finally, the non-minimal coupling constant $\xi$ and the quartic self-coupling constant $\lambda$ can be taken as functions of $\varphi$. The $\varphi$-dependence would mimic their RG-evolution once extra degrees of freedom are included into the theory, that are coupled to the scalar field.

\begin{figure}[t]
\center{\includegraphics[scale=0.6]{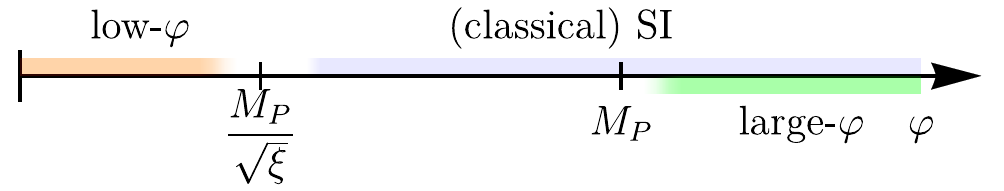}}
\caption{The regions of magnitudes of the scalar field of the model (\ref{L0_J}) with the condition (\ref{CondOnXi}) implemented. }
\label{fig:Scales}
\end{figure}

For the sake of simplicity, we require
\begin{equation}\label{CondOnXi}
\sqrt{\xi}\gg 1 \; ,
\end{equation}
which provides us with a good separation of the Planck and the scale symmetry restoration scales. In what follows, we will refer to the range of magnitudes $\vert\varphi\vert\gg M_P/\sqrt{\xi}$ as the (classically) SI regime of the model, while the sub-range $\vert\varphi\vert\gg M_P$ will be referred to as the large-$\varphi$ regime (see fig. \ref{fig:Scales}).

To study classical configurations arising in the model (\ref{L0_J}), it is convenient to eliminate the non-minimal coupling of the scalar field to gravity. This is achieved by making a Weyl rescaling of the metric
\begin{equation}\label{WeylResc}
\tilde{g}_{\mu\nu}=\Omega^2g_{\mu\nu} \; , ~~~~ \Omega^2=\dfrac{M_P^2+\xi\varphi^2}{M_P^2} \; .
\end{equation}
The Lagrangian of the model is rewritten as\footnote{Transformation of different quantities under the Weyl rescaling can be found, e.g., in \cite{Carneiro:2004rt}. Note that the transformed Ricci scalar contains a total derivative term which is absorbed in the process of transformation of the boundary term and, hence, does not appear in the Lagrangian of the model.}
\begin{equation}\label{L0_E}
\dfrac{\mathcal{L}_{\varphi,g}}{\sqrt{\tilde{g}}}=-\dfrac{1}{2}M_P^2\tilde{R}+\dfrac{1}{2a(\varphi)}(\tilde{\partial} \varphi)^2+\tilde{V}(\varphi) \; ,
\end{equation}
\begin{equation}
a(\varphi)=\dfrac{\Omega^4}{\Omega^2+6\xi^2\varphi^2/M_P^2} \; , ~~~~ \tilde{V}(\varphi)=V(\varphi)\Omega^{-4} \; ,
\end{equation}
where by $(\tilde{\partial}\varphi)^2$ we understand the kinetic term in which the partial derivatives are contracted with the transformed metric $\tilde{g}^{\mu\nu}$.

Let us specify the type of classical configurations suitable for our purposes. First, we restrict the analysis to $O(4)$-symmetric configurations.\footnote{Apart from allowing a simple analytical treatment, the maximally-symmetric case follows naturally from the analysis below, see the footnote on p.5.} To this end, we apply the following Ansatz for the metric,
\begin{equation}\label{MetricAnsatz}
d\tilde{s}^2=f^2(\rho)d\rho^2+\rho^2d\Omega_3^2 \; .
\end{equation}
As we will see shortly, with this somewhat nonstandard choice of the Ansatz, the $00$-component of the Einstein equations is, in fact, an \textit{algebraic} equation for $f$.

Second, we require the configuration to obey the vacuum boundary conditions at infinity\footnote{The first of the conditions in eq. (\ref{CondAtInf}) is corrected if the asymptotic geometry is not flat. The correction is negligible provided that the cosmological length exceeds significantly the characteristic size of the euclidean configuration. This will be true in phenomenological implications of our analysis. Note also that self-consistency requires the scalar field to approach its actual vev in eq. (\ref{CondAtInf}). Again, the difference can be neglected given that the size of the configuration is much smaller than $\langle \varphi\rangle^{-1}$.}
\begin{equation}\label{CondAtInf}
f^2(\rho)\rightarrow 1 \; , ~~~ \varphi(\rho)\rightarrow 0 \; , ~~~ \rho\rightarrow\infty \; .
\end{equation}

To stay close to the actual Higgs physics, we allow the quartic coupling $\lambda$ as a function of $\varphi$ to develop a domain of negative values at some magnitudes of $\varphi$. In this case, the asymptotics specified by eq. (\ref{CondAtInf}) do not approach the true vacuum state of the theory. The model (\ref{L0_E}) then admits the regular solution obeying the condition (\ref{CondAtInf}) --- the bounce \cite{Coleman:1977py,Coleman:1980aw}--- which interpolates between the regions of the true and false vacua. As will become clear later, the bounce is not a configuration suitable for our purposes. 

Apart from the possible bounce, the model (\ref{L0_E}) admits a family of singular configurations that also satisfy the boundary conditions (\ref{CondAtInf}). To find their large-$\varphi$ asymptotics, we write the equations of motion following from the Lagrangian (\ref{L0_E}) in the SI regime and with the Ansatz (\ref{MetricAnsatz}) applied,
\begin{equation}\label{SIEoMs}
\partial_\rho\left(\dfrac{\rho^3\varphi'}{\varphi f}\right)=0 \; , ~~~ f^2=1-\dfrac{\rho^2\varphi'^2}{6a_{SI}\varphi^2} \; , 
\end{equation}
where
\begin{equation}\label{aSI}
a_{SI}=\dfrac{1}{1/\xi+6} \; .
\end{equation}
From this we deduce the behavior of the singular configurations near the origin
\begin{equation}\label{HEInsBehavior}
\varphi\sim \rho^{-\gamma} \; , ~~~ \tilde{R}\sim\rho^{-6} \; , ~~~ \gamma=\sqrt{6a_{SI}} \; , ~~~\rho\rightarrow 0 \; .
\end{equation}
We observe that the physical singularity forms at the center of the configuration. In what follows, we will refer to such configurations as singular shots.\footnote{The configurations of this type were studied before in context with the cosmological initial value problem \cite{Hawking:1998bn,Garriga:1998tm,Vilenkin:1998pp,Turok:1999fe}. In those works, they are referred to as (singular) instantons. In this paper, we would like to reserve the name ``instanton'' for a unique configuration of the family, that obeys an additional boundary condition, see sec. \ref{ssec:Source}. The name ``shot'' is inspired by \cite{Andreassen:2016cvx}.} In proceeding with their studies, one must handle carefully the singular point. The key observation here is that a classical divergence of the field is normally associated with the classical field source acting at the singular point. We start the next section with developing this kind of treatment for the asymptotics (\ref{HEInsBehavior}).

\section{The Mechanism}
\label{sec:Mechanism}

\subsection{Making the instantaneous scalar field source}
\label{ssec:Source}

Our aim is to convert one of the singular shots of the model (\ref{L0_E}) into a valid albeit singular instanton solution by imposing a boundary condition at the singular point. To this end, one must supplement the Lagrangian (\ref{L0_E}) with the source term of the scalar field. This is achieved by exponentiating the field variable $\varphi$ at large magnitudes,\footnote{The condition $\varphi>0$ imposed by eq. (\ref{HiggsExp}) is not restrictive, since the singular shots studied in sec. \ref{sec:Model} are monotonic functions of the radial coordinate with the asymptotics (\ref{CondAtInf}) and (\ref{HEInsBehavior}). }
\begin{equation}\label{HiggsExp}
\varphi\rightarrow M_Pe^{\bar{\varphi}/M_P} \; .
\end{equation}
This nonlinear change of the field variable arises naturally when one rewrites the original Lagrangian of the model, given in eq. (\ref{L0_J}), in the form (\ref{L0_E}). Indeed, from eq. (\ref{L0_E}) we see that $\bar{\varphi}$ is, in fact, a canonical variable for the scalar field in the limit $\varphi\gg M_P/\sqrt{\xi}$.\footnote{Up to a coefficient of the order of one, according to eq. (\ref{CondOnXi}).} Therefore, we can endow eq. (\ref{HiggsExp}) with the physical meaning by saying that it is $\bar{\varphi}$ that carries the valid degree of freedom of the scalar field at large magnitudes of the latter. Here, one can notice an analogy with the gauge theories, in the confinement phase of which the description must be performed in terms of the Wilson loops, not the gauge field itself \cite{Polyakov:1978vu}. 

\begin{figure*}[t]
\begin{center}
\begin{minipage}[h]{0.49\linewidth}
\center{\includegraphics[scale=0.65]{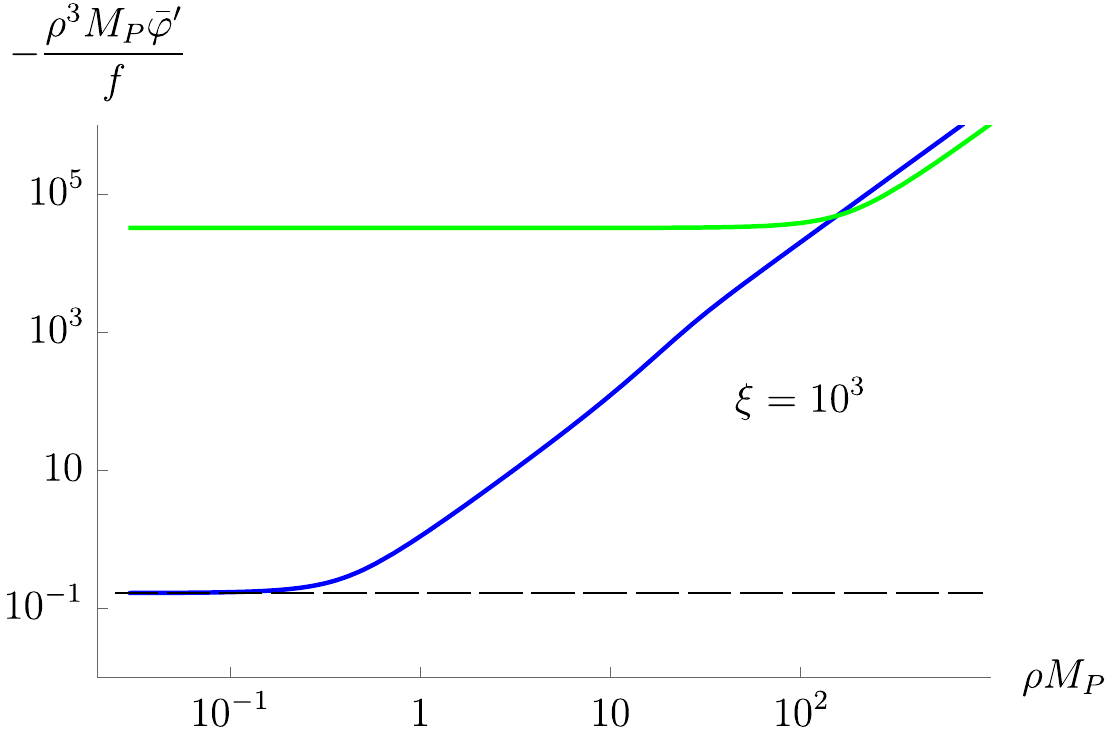}}
\end{minipage}
\begin{minipage}[h]{0.49\linewidth}
\center{\includegraphics[scale=0.65]{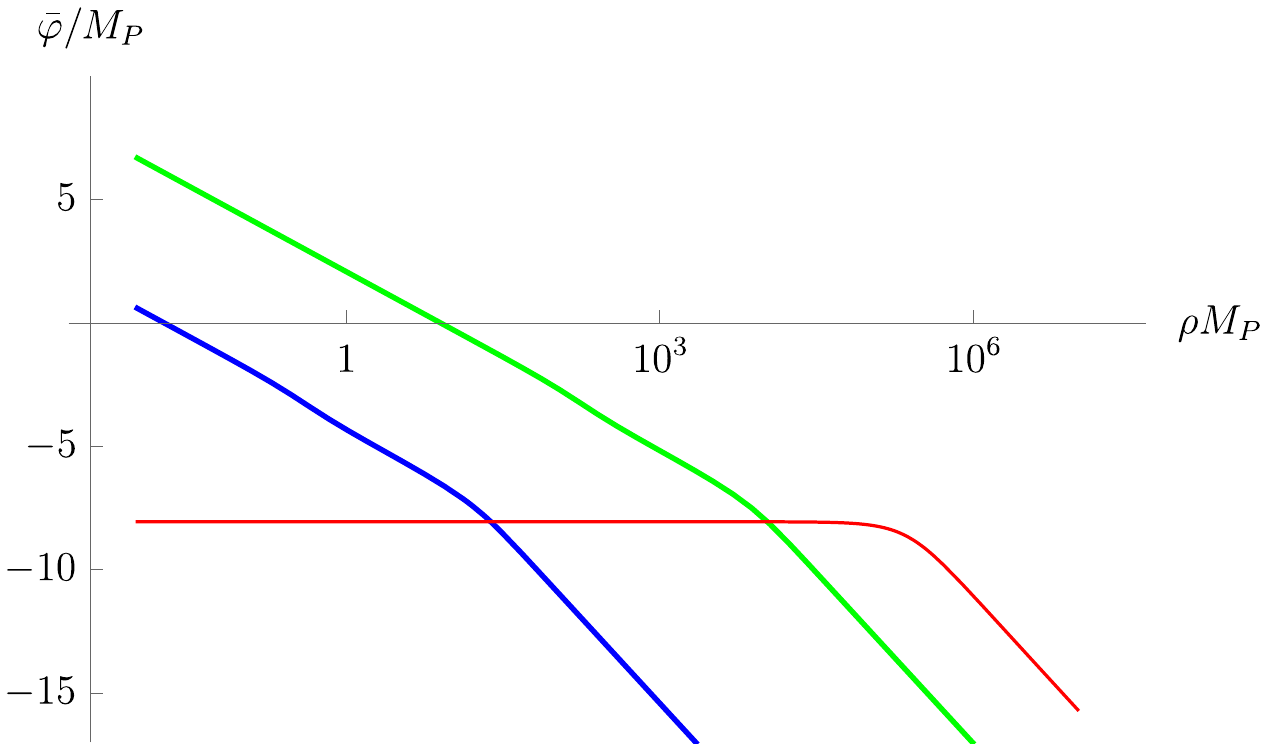}}
\end{minipage}
\caption{Two featured singular configurations of the model (\ref{L0_E}). The shot painted blue matches the scalar field source in the r.h.s. of eq. (\ref{CorrectAsymptotics}), hence it is a valid singular instanton. The shot painted green is the one with the large euclidean action; for illustration, we choose for it $\bar{S}_E=40$. We take the potential for the scalar field coinciding with the (RG-improved) Higgs potential in the SM with the central values of the top quark mass $m_t=172.25$ GeV \cite{Castro:2017yxe}, the Higgs mass $m_H=125.09$ GeV \cite{Aad:2015zhl}, and the field-dependent normalization point $\mu=\bar{\varphi}$. The left panel shows the short-distance asymptotics of the relevant combination of the fields, the dashed line marks the value $(6+1/\xi)^{-1}$. The right panel shows the behavior of $\bar{\varphi}$ as the singularity is approached. For illustrative purposes, the bounce is also plotted in red.}
\label{Fig:SingShots1}
\end{center}
\end{figure*}

The field redefinition (\ref{HiggsExp}) results in the appearance of the desired source term in the process of evaluation of the vev $\langle\varphi\rangle$. Indeed, the corresponding part of the path integral in eq. (\ref{HiggsVEV}) becomes, schematically,
\begin{equation}
\int_{\varphi\gtrsim M_P/\sqrt{\xi}} \mathcal{D}\varphi \varphi e^{-S_E} \rightarrow M_P\int_{\bar{\varphi}\gtrsim M_P\log(1/\sqrt{\xi})}\mathcal{D}\bar{\varphi} Je^{-W} \; ,
\end{equation}
\begin{equation}\label{W}
W=-\bar{\varphi}(0)/M_P+S_E
\end{equation}
with $J$ the corresponding Jacobian, and we used the translational invariance of the theory to put the instantaneous source of the scalar field at the origin of coordinates. We now consider $W$ as a functional whose saddle points are to be studied. Introducing the radial delta-function $\delta(\rho)$ such that $\bar{\varphi}(0)=\int d\rho\delta(\rho)\bar{\varphi}(\rho)$, we find the modification of eqs. (\ref{SIEoMs}) in the large-$\varphi$ (or, equivalently, large-$\bar{\varphi}$) regime,
\begin{equation}\label{HEEoMs}
\partial_\rho\left(\dfrac{\rho^3\bar{\varphi}'}{fa_{SI}}\right)=-\dfrac{1}{M_P}\delta(\rho) \; , ~~~ f^2=1-\dfrac{\rho^2\bar{\varphi}'^2}{6a_{SI}M_P^2} \; ,
\end{equation}
where $a_{SI}$ is given in eq. (\ref{aSI}), and the asymptotic behavior of the scalar field is now
\begin{equation}\label{HEasympt}
\bar{\varphi}= -M_P\sqrt{6a_{SI}}\log\rho M_P+C \; , ~~~ \rho\rightarrow 0
\end{equation}
with $C$ a constant used to match with the asymptotics (\ref{CondAtInf}) at large $\rho$. We observe that the exponentiation of the $\varphi$-variable leads to the fixation of the position of the center of the singular shots and makes one of them the legitimate solution of the variational problem $\delta W/\delta\bar{\varphi}=0$ with the boundary condition (\ref{CondAtInf}).\footnote{Note also that introducing a point source prefers maximally-symmetric configurations seeded around it. Should the less symmetric configurations exist, we assume that their contribution to the path integral is suppressed compared to the $O(4)$-symmetric case, cf. footnote on p.4.} In what follows, we will refer to this solution as the singular instanton. To simplify the consideration, in the rest of the paper we will work with the $\bar{\varphi}$-variable in the entire range of magnitudes, bearing in mind that, by construction, it carries the valid degree of freedom only at $\bar{\varphi}\gtrsim M_P\log(1/\sqrt{\xi})$. We would like to use the singular instanton of this type as a saddle point of the functional $W$, that contributes to the vev $\langle\varphi\rangle$. In the saddle-point approximation (SPA), this amounts to saying that
\begin{equation}\label{HiggsVEVSPA}
\langle\varphi\rangle\approx M_P e^{-\bar{W}} \; ,
\end{equation}
where $\bar{W}$ is the value of $W$ computed on the instanton. 

Formula (\ref{HiggsVEVSPA}) manifests the appearance of a new scale in the model (\ref{L0_E}). We are interested in the case when this scale is much smaller than the original scale $M_P$ (or $M_P/\sqrt{\xi}$). For this to happen, one should require
\begin{equation}\label{CondOnW}
\bar{W}\gg 1 \; .
\end{equation}
Of course, if in a particular situation one reveals that $\bar{W}$ is nearly zero or negative, the SPA is not applicable, and eq. (\ref{HiggsVEVSPA}) is not valid. The possible interpretation of this case is that the non-perturbative effects of quantum gravity are strong, and, hence, no new scale appears. If, on the other hand, eq. (\ref{CondOnW}) is fulfilled, these effects are suppressed, and the hierarchy of scales is generated. Our goal for the rest of this section is to find when it is possible to satisfy eq. (\ref{CondOnW}) in the model (\ref{L0_E}) or its modifications.

\subsection{Attempting to compute the vev in the simple model}
\label{ssec:Attempt}

In trying to compute $\bar{W}$ in the model (\ref{L0_E}), one immediately encounters multiple issues. We describe them here, and in sec. \ref{ssec:Der} and \ref{ssec:Poly} the large-$\bar{\varphi}$ modifications of the model are studied with the aim to cure them.

\begin{figure*}[t]
\begin{center}
\begin{minipage}[h]{0.49\linewidth}
\center{\includegraphics[scale=0.65]{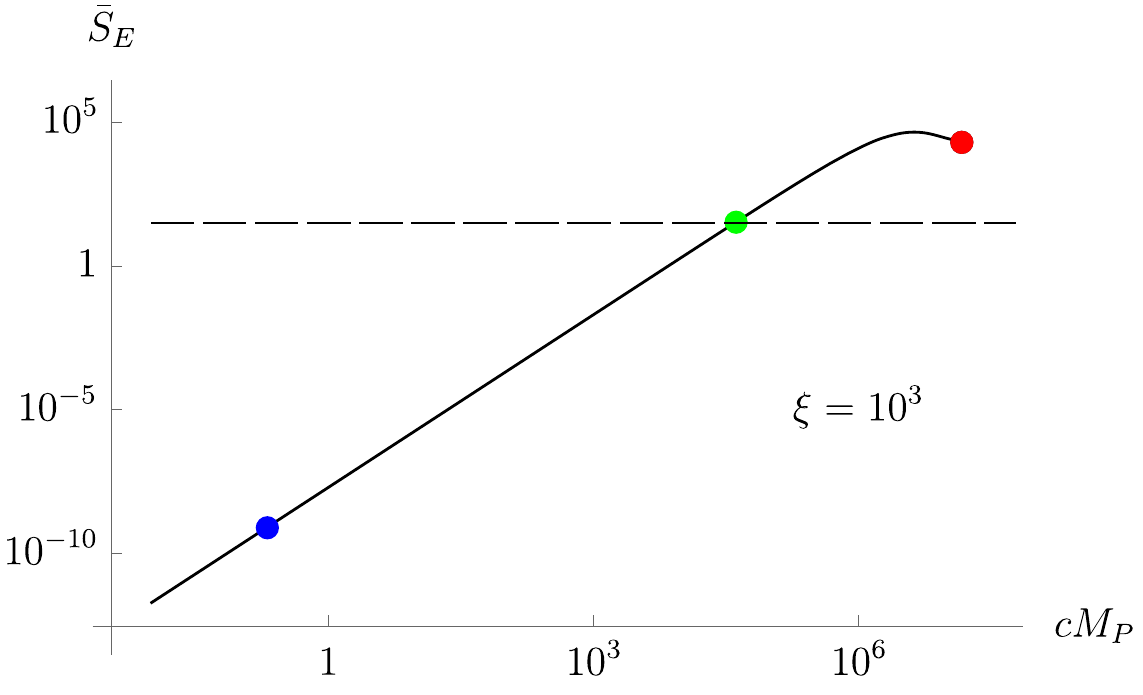}}
\end{minipage}
\begin{minipage}[h]{0.49\linewidth}
\center{\includegraphics[scale=0.65]{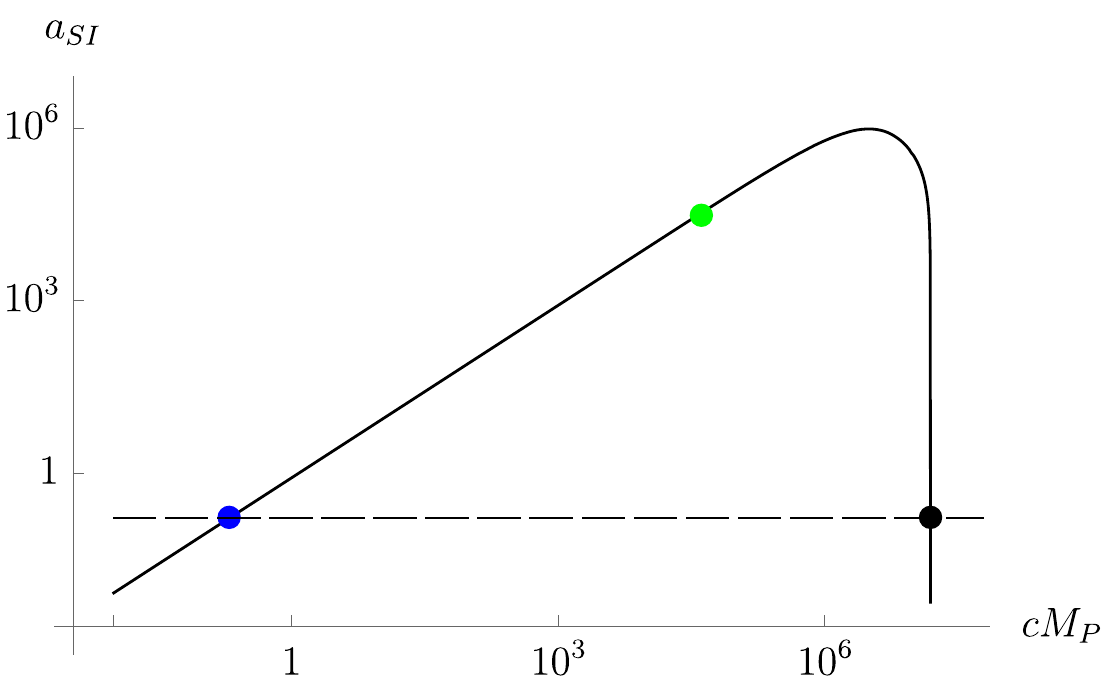}}
\end{minipage}
\caption{The family of singular configurations of the model (\ref{L0_E}), parametrized by their fall-off at infinity, $\varphi=c\rho^{-2}$, $\rho\rightarrow\infty$. We take the same potential for the scalar field as in fig. \ref{Fig:SingShots1}. The colored circles correspond to the shots shown in fig. \ref{Fig:SingShots1}. The left panel shows the euclidean action of the shot plotted against the parameter $c$, the dashed line corresponds to $\bar{S}_E=40$. The right panel shows the value of the scalar field source necessary to match with the short-distance asymptotics of the solution, the dashed line corresponds to the value $(6+1/\xi)^{-1}$ (see eq. (\ref{CorrectAsymptotics})). The black circle indicates another singular instanton with the correct asymptotic behaviour; contrary to other configurations, this instanton lies close to the bounce everywhere except the core, and its euclidean action is nearly the same as the one of the bounce. In the SPA, one should exclude this instanton from consideration in favour of the one with the small euclidean action.}
\label{Fig:SingShots2}
\end{center}
\end{figure*}

(\textit{i}) It is immediately seen from eq. (\ref{HEasympt}) that $\bar{\varphi}(0)=\infty$, hence $\bar{W}$ is divergent. In order to extract a meaningful information about the contribution of the singular instanton to the vev $\langle\varphi\rangle$, an accurate treatment of this divergence is required.

(\textit{ii}) We did not present the semiclassical parameter that would justify the SPA made in going from eq. (\ref{HiggsVEV}) to eq. (\ref{HiggsVEVSPA}). However, such justification can be made \textit{a posteriori} provided that we have an explicit solution.

(\textit{iii}) Let us compute the euclidean action of the singular instanton $\bar{S}_E$ in the model (\ref{L0_E}). Making use of the Einstein equations, one obtains
\begin{equation}\label{OnShellAction}
\bar{S}_E=-\int d^4x\sqrt{\tilde{g}}\tilde{V}(\bar{\varphi}) \; .
\end{equation}
If one leaves aside for the moment the issue with the singular term in $W$, then $\bar{S}_E$ is to provide the desired suppression of the Planck scale in eq. (\ref{HiggsVEVSPA}). This would imply
\begin{equation}\label{S0}
\bar{S}_E\gg 1 \; .
\end{equation}
For $\bar{S}_E$ to be positive, the field-dependent coupling constant $\lambda$ must be negative at some values of $\bar{\varphi}$, thus admitting the bounce solution alongside with the instanton solution. Bearing in mind phenomenological applications of our analysis, we require the euclidean action of the bounce to be large enough in order to ensure the sufficient long-liveness of the false vacuum. With this condition and the condition (\ref{CondOnXi}) implemented, we investigate numerically the singular shots obeying the asymptotics (\ref{CondAtInf}) and, in particular, the singular instanton for which
\begin{equation}\label{CorrectAsymptotics}
\dfrac{1}{a_{SI}}\dfrac{\rho^3\bar{\varphi}'}{f}\rightarrow-\dfrac{1}{M_P} \; , ~~~ \rho\rightarrow 0 \; .
\end{equation}

The results of the analysis are presented in fig. (\ref{Fig:SingShots1}) and (\ref{Fig:SingShots2}). We take $\xi=10^3$, and the $\bar{\varphi}$-dependence of $\lambda$ as if it underwent the RG-running in the SM with the central values of the parameters of the latter.\footnote{Here and below we assume that the RG-running of the non-minimal coupling $\xi$ in the theory can be neglected. This is justified by noticing that $\xi$ evolves rather slowly with the energy scale increasing, provided that it is always far from the conformal limit. Moreover, the results of our analysis will eventually be insensitive to the particular value of $\xi$.} We observe that the requirement of the correct asymptotic behavior in the large-$\bar{\varphi}$ limit is in a sharp contrast with the requirement to have the large value of $\bar{S}_E$. In fact, the incompatibility of the two conditions cannot be overcome regardless the shape of the potential for the scalar field. The reason is that the singular instanton turns out to be insensitive to the details of the potential at low magnitudes of $\bar{\varphi}$, since it shoots too fast through this region of magnitudes, and no substantial contribution to the euclidean action can be produced. This forces us to conclude that in the model (\ref{L0_E}) it is impossible to make the singular shot with the conditions (\ref{S0}) and (\ref{CorrectAsymptotics}) both satisfied. The euclidean action of the singular instanton turns out to be nearly zero, the SPA is not applicable, and the non-perturbative effects are expected to drive the value of $\langle\varphi\rangle$ close to the Planck scale.

Problems $(i)-(iii)$ pose serious obstacles to our analysis. The possible way out is to modify the model (\ref{L0_E}) in the large-$\bar{\varphi}$ regime. The necessity for such modification comes naturally once we notice that at $\xi\varphi^2\ll M_P^2$ the Lagrangian (\ref{L0_E}), in fact, describes the low energy limit of the theory we are ultimately interested in. Hence, considering different higher-dimensional operators supplementing the model (\ref{L0_E}) at large values of $\bar{\varphi}$ and its derivative is in agreement with the strategy of probing different possible UV properties of the theory reducing to the CI SM and GR at low energy scales, of which proper UV completion we are not aware. The applicability limit of the low energy description is determined by considering unitarity bounds of $n$-particles scattering amplitudes on top of the background vacuum configuration. The UV cutoff $\Lambda$ of the model (\ref{L0_E}), determined in this way, is found to be \cite{Burgess:2009ea,Barbon:2009ya,Bezrukov:2010jz}
\begin{equation}
\Lambda\sim M_P/\xi \; .
\end{equation}
This justifies the usage of the Planck-suppressed in the low-$\bar{\varphi}$ limit operators which are composed of the scalar and the metric fields. Below we consider some particular classes of such operators in an attempt to improve the large-$\bar{\varphi}$ properties of the singular instanton and, eventually, to cure problems $(i)-(iii)$.

\subsection{Shaping the large field limit with derivative operators}
\label{ssec:Der}

We would like to study how the short-distance behavior of the singular shots of the model of sec. \ref{sec:Model} is changed when new operators are added into the model. We start by introducing an operator containing the higher power of the derivative of the scalar field,
\begin{equation}\label{O_n_J}
\mathcal{O}_n=\sqrt{g}\;\delta_n\dfrac{(\d\varphi)^{2n}}{(M_P\Omega)^{4n-4}}
\end{equation} 
with $\Omega$ given in eq. (\ref{WeylResc}) and $\delta_n$ a dimensionless constant. The operator (\ref{O_n_J}) is suppressed by $M_P^{4n-4}/\delta_n$ in the low energy low-$\varphi$ limit and becomes independent of $M_P$ in the large-$\varphi$ limit. For simplicity, here we limit the consideration to a single $(n=2)$-operator. The general case will be commented on in sec. \ref{sec:DiscConcl}. Making the Weyl rescaling of the metric (\ref{WeylResc}) and changing the variable according to eq. (\ref{HiggsExp}), we obtain the modification of the model in the SI regime by the operator
\begin{equation}\label{O_2_E}
\tilde{\mathcal{O}}_2=\sqrt{\tilde{g}}\;\delta\dfrac{(\tilde{\d}\bar{\varphi})^4}{M_P^{4}} \;, ~~~ \delta=\delta_2/\xi^2 \; ,
\end{equation}
and we assume $\delta\lesssim 1$. Note that the variation of the operator (\ref{O_2_E}) with respect to $\bar{\varphi}$ is a total derivative; this simplifies significantly the analytical treatment of the singular shots in the model (\ref{L0_E})+(\ref{O_2_E}). Applying the Ansatz (\ref{MetricAnsatz}), we obtain the modified equation of motion for the scalar field in the SI regime,
\begin{equation}\label{EqO}
\dfrac{1}{a_{SI}}\dfrac{\rho^3\bar{\varphi}'}{f}+\dfrac{4\delta}{M_P^4}\dfrac{\rho^3\bar{\varphi}'^3}{f^3}=-\dfrac{1}{M_P} \; .
\end{equation}
Denote by $\bar{\rho}$ the size of the region in which the second term in the l.h.s. of eq. (\ref{EqO}) is dominant. The asymptotic behavior of the singular instanton in this region is given by
\begin{equation}\label{HEAsumptotics}
\bar{\varphi}'\sim M_P^2\delta^{-1/6} \; , ~~~ f\sim \rho M_P\delta^{1/6} \; , ~~~ \rho\lesssim\bar{\rho} \; .
\end{equation}
First of all, we observe that $\bar{\varphi}$ is not divergent at the origin any more. Hence, the magnitude of the scalar field at the center of the instanton becomes finite. This cures problem \textit{(i)} of sec. \ref{ssec:Attempt}. Note that, despite being finite, the instanton remains to be singular; in particular, the scalar curvature $\tilde{R}\sim\rho^{-2}$, $\rho\rightarrow 0$. Hence, the instantaneous source of the field $\bar{\varphi}$ is still necessary in obtaining the solution. Next, we notice that the core region $\rho\lesssim\bar{\rho}$ of the instanton provides a finite contribution to the euclidean action. Indeed, the latter becomes
\begin{equation}
\bar{S}_E=\int d\rho\:\bar{\L} \; , ~~~ \bar{\L}=2\pi^2\rho^3f\left[\delta\dfrac{\bar{\varphi}'^4}{f^4M_P^4}-\tilde{V}(\bar{\varphi})\right] \; ,
\end{equation}
where $\bar{\L}$ is the Lagrangian of the model (\ref{L0_E})+(\ref{O_2_E}), computed on the singular instanton. In the large-$\bar{\varphi}$ regime, it is given by
\begin{equation}\label{L_E_d}
\left.\bar{\L}\right\vert_{\rho\lesssim\bar{\rho}}=2^{-4/3}M_P\delta^{-1/6} \; .
\end{equation}
Finally, the total derivative form of the scalar field equation of motion, given in eq. (\ref{EqO}), implies that the short-distance asymptotics of the solution (\ref{CorrectAsymptotics}), achieved in the SI sub-region well below $M_P$, remains unchanged once the operator (\ref{O_2_E}) becomes important. Hence, according to the discussion in sec. \ref{ssec:Attempt}, the low-$\bar{\varphi}$ part of the instanton cannot provide a suitable contribution to the euclidean action. As for the large-$\bar{\varphi}$ part, one would expect its contribution to be tunable by the parameter $\delta$. However, from eqs. (\ref{HEasympt}), (\ref{EqO}) and (\ref{HEAsumptotics}) it follows that
\begin{equation}\label{RhoBar}
\bar{\rho}\sim M_P^{-1}\delta^{1/6}a^{1/2}_{SI} \; .
\end{equation}
From this and eq. (\ref{L_E_d}) one now sees that $\bar{S}_E$ does not experience any power-like dependence on $\delta$.\footnote{This fact remains true for any operator of the form (\ref{O_n_J}) inserted into Lagrangian (\ref{L0_J}), as well as for any analytic function summable from the series of such operators.} Hence, for the model (\ref{L0_E})+(\ref{O_2_E}) one expects again $\langle\varphi\rangle$ to lie close to the Planck scale, and the question of how to generate the large value of $W$ persists.

\subsection{Making the hierarchy of scales with polynomial operators}
\label{ssec:Poly}

Knowing the asymptotics of the scalar field in the large-$\bar{\varphi}$ and low-$\bar{\varphi}$ regions of the instanton, one can make a rude estimation of its magnitude at the center of the instanton. In the model (\ref{L0_E})+(\ref{O_2_E}) it is found to be
\begin{equation}
\bar{\varphi}(0)/M_P\sim a_{SI}^{1/2}(\log\delta-3\log a_{SI}+\mathcal{O}(1)) \; .
\end{equation}
\begin{figure}[h]
\center{\includegraphics[scale=0.65]{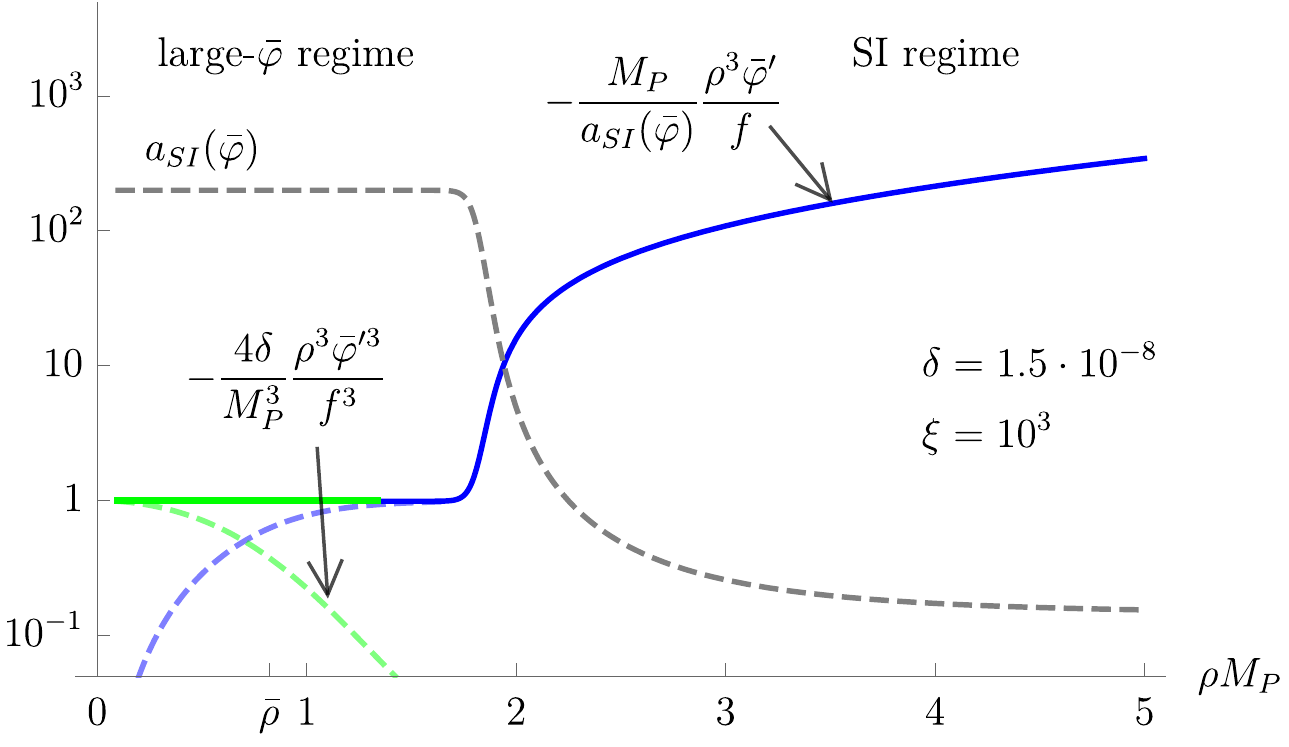}}
\caption{The l.h.s. of eq. (\ref{EqO}) in the units of $M_P^{-1}$ and with $a_{SI}$ given in eq. (\ref{a_HE}). The solid line represents the sum of the two terms; it approaches $M_P^{-1}$ but does not coincide with it unless $a_{SI}$ freezes. The colored dashed lines show the contributions from each of the terms. The potential for the scalar field and the value of $\xi$ are the same as in fig. \ref{Fig:SingShots1}. }
\label{Fig:Solution}
\end{figure}
\begin{figure*}[t]
\begin{center}
\begin{minipage}[h]{0.49\linewidth}
\center{\includegraphics[scale=0.65]{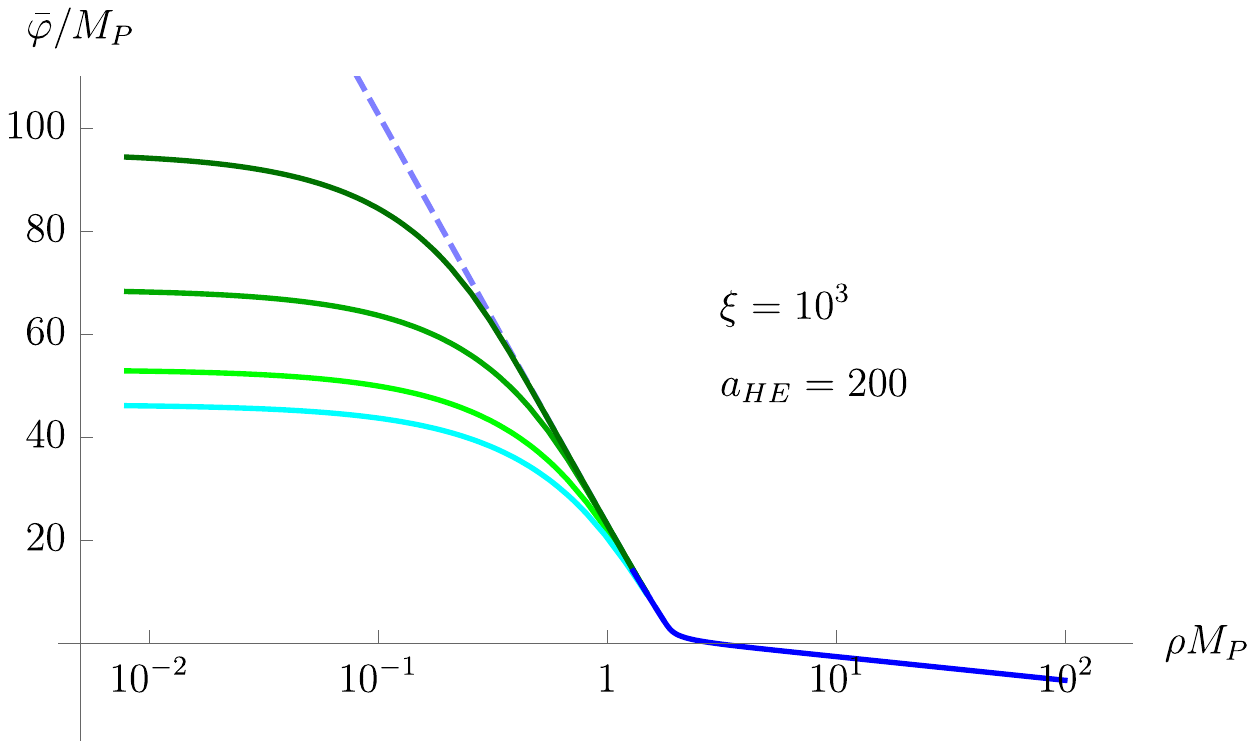}}
\end{minipage}
\begin{minipage}[h]{0.49\linewidth}
\center{\includegraphics[scale=0.65]{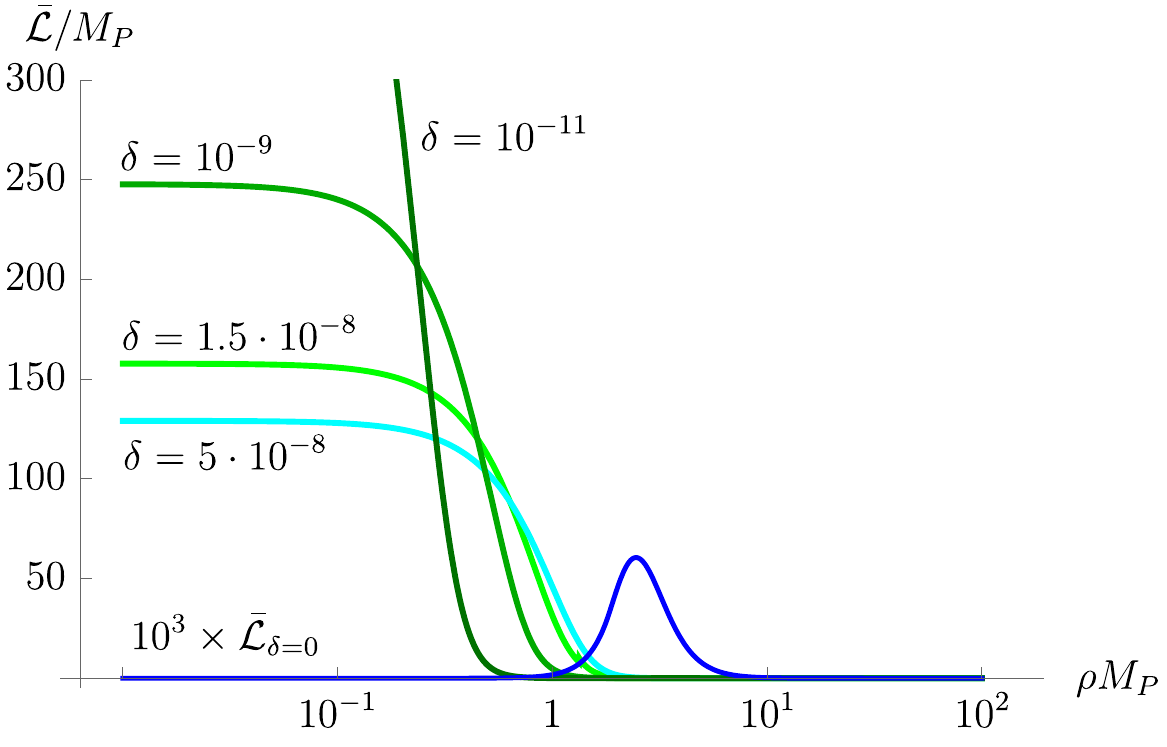}}
\end{minipage}
\caption{The family of singular instantons in the model specified by eqs. (\ref{L_J_Poly}), (\ref{F,G}), to which the Weyl rescaling (\ref{WeylResc}) is applied. The left panel demonstrates the finite short-distance asymptotics of the instantons, the dashed line shows the case $\delta=0$. The right panel shows the corresponding Lagrangians. We observe an agreement with eqs. (\ref{L_E_d}) and (\ref{RhoBar}). One also sees that the sizeable contribution to the euclidean action comes from the large-$\bar{\varphi}$ region. The potential for the scalar field and the value of $\xi$ are the same as in fig. \ref{Fig:SingShots1}.}
\label{Fig:VarDelta}
\end{center}
\end{figure*}
From this and eqs. (\ref{L_E_d}), (\ref{RhoBar}), we deduce the power-like dependence of $\bar{W}$ on $a_{SI}$,\footnote{We made use of the fact that the contribution of the singular instanton to $W$ outside the large-$\bar{\varphi}$ region is negligible. In what follows, this will remain true.}
\begin{equation}\label{WOnShell}
\bar{W}\sim a_{SI}^{1/2} \; .
\end{equation}
Hence, one can expect that the large values of $W$ can be achieved by tuning the value of $a_{SI}$. However, from eq. (\ref{aSI}) we see that $a_{SI}$ is confined in the region
\begin{equation}\label{CondOnaCI}
0<a_{SI}<1/6 \; ,
\end{equation}
which makes it impossible to fulfill relation (\ref{CondOnW}). Hence, no hierarchy is generated in the model (\ref{L0_E})+(\ref{O_2_E}). The possible resolution of this issue is to look for further modifications of the model in the large-$\bar{\varphi}$ regime, that would lead to the modification of the allowable range of values of $a_{SI}$. To this end, consider the following Lagrangian,
\begin{align}\label{L_J_Poly}
\dfrac{\mathcal{L}_{\varphi,g}}{\sqrt{g}}=-\dfrac{M_P^2}{2}F(\varphi/M_P)R & +\dfrac{1}{2}G(\varphi/M_P)(\partial\varphi)^2  \nonumber \\  
& + \delta\xi^2\dfrac{(\partial\varphi)^4}{(M_P\Omega)^4}+\dfrac{\lambda}{4}\varphi^4 \; ,
\end{align}
where $F$ and $G$ are rational functions of $\varphi/M_P$ taken so as to reproduce the Lagrangian (\ref{L0_J}) in the low-$\varphi$ limit, and $\Omega$ is given in eq. (\ref{WeylResc}). The simplest, but not unique, possibility leading to the desired change of the range of $a_{SI}$, is to choose
\begin{equation}\label{F,G}
F=1+\xi \varphi^2/M_P^2 \; , ~~~ G=\dfrac{1+\kappa \varphi^2/M_P^2}{1+\varphi^2/M_P^2} 
\end{equation}
with $\kappa$ some constant. Then, one can show that the coefficient $a_{SI}$ becomes field-dependent,
\begin{equation}\label{a_HE}
a_{SI}=\dfrac{1}{\alpha/\xi+6} \; ,
\end{equation}
where, in terms of the field variable $\bar{\varphi}$,
\begin{equation}
\alpha=\dfrac{1}{2}(1-\tanh(\bar{\varphi}/M_P))+\dfrac{\kappa}{2}(1+\tanh(\bar{\varphi}/M_P)) \; .
\end{equation}
Hence, the asymptotic value of $a_{SI}$ in the large-$\bar{\varphi}$ regime modifies to
\begin{equation}\label{aHE}
a_{SI}\rightarrow a_{HE}=\dfrac{1}{\kappa/\xi+6} \; , ~~~ \rho\rightarrow 0 \; , ~~~ \bar{h}\gtrsim M_P \; ,
\end{equation}
as compared with eq. (\ref{aSI}). By tuning $\kappa$, one can make $a_{HE}$ as large as necessary, thus ``enhancing'' the strength of the scalar field source by a suitable amount. Finally, eq. (\ref{WOnShell}) becomes
\begin{equation}\label{WHEOnShell}
\bar{W}\sim a_{HE}^{1/2} \; .
\end{equation}

Let us now study the singular instantons arising in the model specified by eqs. (\ref{L_J_Poly}), (\ref{F,G}), to which the Weyl rescaling (\ref{WeylResc}) is applied. For simplicity, we assume that the transition between the low-$\bar{\varphi}$ and the large-$\bar{\varphi}$ values of $a_{SI}$ occurs before the asymptotics of the instanton becomes dominated by the quartic derivative term (\ref{O_2_E}). This provides us with a separation of regions at which the quartic derivative operator and the polynomial operators start affecting the behavior of the solution. According to eq. (\ref{RhoBar}), the requirement of such separation puts an upper bound on $\delta$,
\begin{equation}\label{CondOnDelta}
a_{HE}^{1/2}\delta^{1/6}\ll 1 \; ,
\end{equation}
which can easily be satisfied in our analysis. Overall, we look for a classical configuration obeying the asymptotics (\ref{CondAtInf}) at large distances, and eq. (\ref{EqO}) with $a_{SI}$ replaced by $a_{HE}$ according to eq. (\ref{aHE}) --- at short distances.

Bearing in mind the insensitivity of the singular shot to the details of the model below the large-$\bar{\varphi}$ regime, which was discussed in sec. \ref{ssec:Attempt}, we focus on the variation of the large-$\bar{\varphi}$ parameters $a_{HE}$ and $\delta$. Then, numerics shows that it is possible for a fixed value of $\delta$ to choose $a_{HE}$ so that relation (\ref{CondOnW}) is satisfied. An example of the solution is presented in fig. \ref{Fig:Solution}. As expected from the discussion in sec. \ref{ssec:Der}, the variation of $\delta$ does not change the picture qualitatively. Fig. \ref{Fig:VarDelta} exemplifies the difference in the large-$\bar{\varphi}$ behaviour of the singular instanton and in the shape of the Lagrangian computed on it, as $\delta$ varies.

\begin{figure}[t]
\center{\includegraphics[scale=0.6]{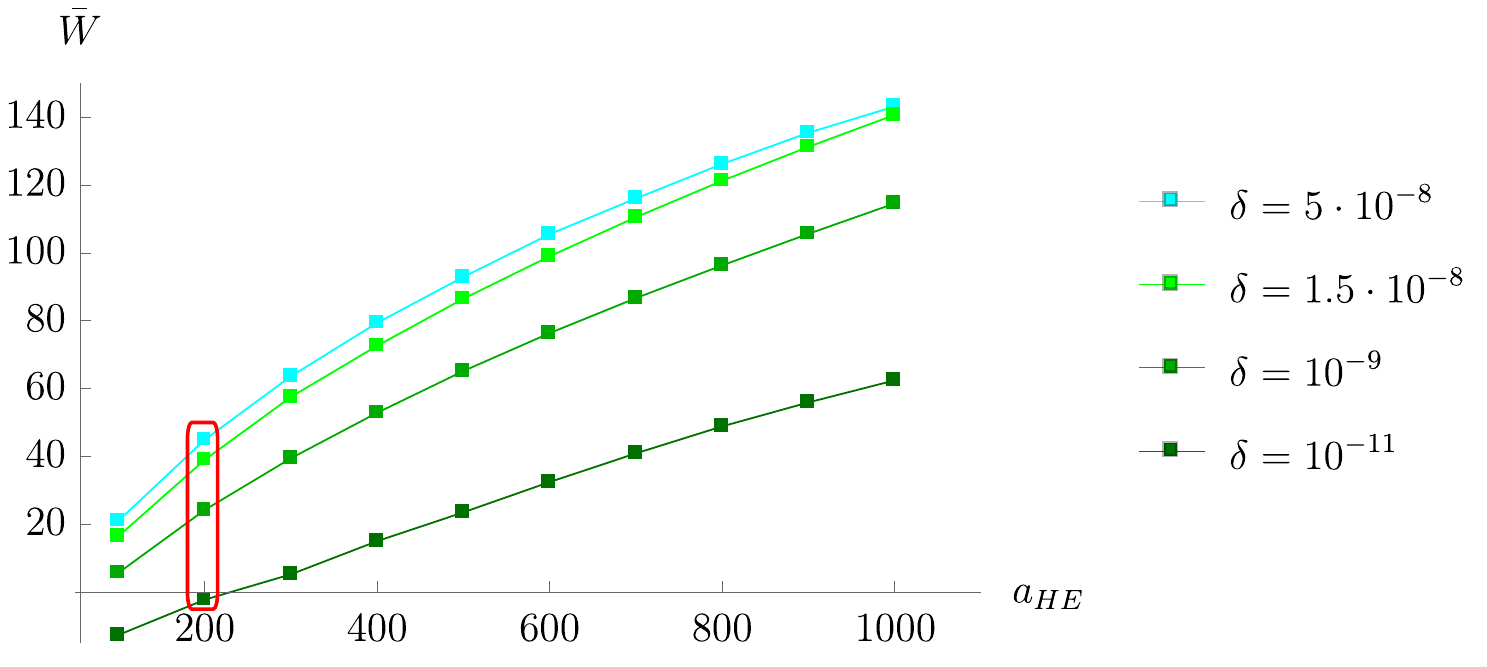}}
\caption{The instanton value of the functional $W$ plotted against the coefficient $a_{HE}$ and with the different choices of the parameter $\delta$. One observes the ambiguity in the choice of $a_{HE}$, $\delta$ leading to a given $\bar{W}$. The red frame marks the solutions studied in fig. \ref{Fig:VarDelta}. The small deviations from the power-law behaviour (\ref{WHEOnShell}) are due to a sub-dominant dependence on $a_{SI}$ and an imperfect separation of the region where $a_{SI}$ varies from the region where the quartic derivative term dominates (see eq. (\ref{CondOnDelta})).}
\label{Fig:Degeneracy}
\end{figure}

It is important to note that the euclidean action in the functional $W$ is saturated in the large-$\bar{\varphi}$ domain, $\bar{\varphi}\gtrsim M_P$, and the contribution from the region of lower magnitudes of the scalar field is completely negligible. This is clearly seen from the right side of fig. \ref{Fig:VarDelta}, which shows the Lagrangian of the model as a function of the distance from the core of the instanton and at different values of $\delta$. We conclude that the power-like estimation for the suppression exponent (\ref{WHEOnShell}) is valid for our solutions; this is checked explicitly in fig. \ref{Fig:Degeneracy}, where $\bar{W}$ is plotted against $a_{HE}$. Furthermore, $a_{HE}^{-1/2}$ can be seen as a small parameter whose appearance as a common multiplier in $\bar{W}$ justifies the SPA made in obtaining eq. (\ref{HiggsVEVSPA}). This observation resolves issue $(ii)$ of sec. \ref{ssec:Attempt}, thus completing the analysis.

\subsection{Summary}
\label{ssec:Summary}

The model of sec. \ref{ssec:Poly} is the one in which we have found the singular instanton with the large value of $W$. The reason we have provided the detailed exposition of other models is that we wanted to make clear the essential ingredients of the mechanism. It is their step-by-step implementation that guided us from the basic model (\ref{L0_J}) to the one described by eqs. (\ref{L_J_Poly}), (\ref{F,G}). We list these ingredients in this short summary, while postponing a more general discussion to sec. \ref{sec:DiscConcl}.
\begin{itemize}
\item The non-minimal coupling of the scalar field to gravity, regulated by the parameter $\xi$. This coupling allowed us to change the variable according to eq. (\ref{HiggsExp}) and, eventually, to introduce the point source of the scalar field. In turn, the source provided us with an additional boundary condition selecting a unique solution from the family of singular configurations.
\item The higher-dimensional operators of the form (\ref{O_n_J}), which regularize the otherwise divergent solution. Adding such operators allowed us to avoid dealing with the divergence of the instanton at the source point.
\item The source enhancement in the large field limit, performed by using the polynomial operators of the form (\ref{F,G}). They enabled us to make a large value of $W$.
\end{itemize}
Regarding the last point, it is interesting to note that in the limit $a_{HE}\rightarrow\infty$, $\delta\rightarrow 0$ the model of sec. \ref{ssec:Poly} acquires the Weyl symmetry in the large field region. To see this, we rewrite the Lagrangian of the model in the SI regime as follows,
\begin{equation}
\dfrac{\L_{\varphi,g,SI}}{\sqrt{g}}=\dfrac{1}{2}\dfrac{1}{6-a_{HE}^{-1}}\varphi^2 R+\dfrac{1}{2}(\d\varphi)^2+\delta\dfrac{(\d\varphi)^4}{\varphi^4} \; .
\end{equation}
As $a_{HE}$ grows and $\delta$ decreases, the Lagrangian approaches a Weyl-invariant point. Thus, in generating the hierarchy of scales, one can make use of the (approximate) asymptotic Weyl symmetry of the theory. As fig. \ref{Fig:Degeneracy} demonstrates, the rate of suppression of the original scale $M_P$ can, in fact, be arbitrarily large.

\section{Implications for the Hierarchy problem}
\label{sec:SM}

The results of the previous section are applied directly to the EW hierarchy problem. We consider the Lagrangian of sec. \ref{ssec:Poly} as describing the Higgs-gravity sector of the theory under consideration. The real scalar field $\varphi$ is identified with the Higgs field degree of freedom in the unitary gauge,
\begin{equation}
\phi=1/\sqrt{2}\;(0,\varphi)^T \; .
\end{equation}
The Higgs-gravity Lagrangian is supplemented with the rest of the low energy content of the theory. One can choose the latter to be that of the SM. All extra fields entering the theory are taken at their vacuum values. The fluctuations of the fields affect the prefactor which in eq. (\ref{HiggsVEVSPA}) is chosen to be approximately equal to $M_P$. The validity of the SPA enables us to believe that the higher-order corrections to eq. (\ref{HiggsVEVSPA}) do not change drastically the leading-order calculation. Furthermore, fig. \ref{Fig:Degeneracy} assures that any corrections coming from the prefactor can be compensated by readjusting the parameters of the theory. According to eq. (\ref{HiggsVEVSPA}), the observed ratio of the Fermi scale to the Planck scale is reproduced when
\begin{equation}
\bar{W}=\log(M_P/v)\approx 37 \; .
\end{equation}
The singular instanton with this value of $W$ is studied in fig. \ref{Fig:Solution}. 

One may wonder how the modifications brought to the Higgs-gravity sector by the derivative and polynomial operators affect the dynamics of the SM fields coupled to the Higgs field. The worrisome observation here is that the coefficient $\alpha$ in eq. (\ref{a_HE}) appears in the Higgs field kinetic term of the lowest order after the Weyl rescaling of the metric is performed. If $a_{HE}$ is demanded to be large enough for the mechanism to work, $\alpha$ becomes negative. When supplementing the Higgs-gravity Lagrangian with the rest of the SM fields, one replaces in the Higgs kinetic term
\begin{equation}
\partial_\mu\rightarrow\mathcal{D}_\mu \; .
\end{equation}
This endangers the dynamics of the gauge fields, as the latter become tachyonic as soon as they interact with the Higgs field through the SM coupling terms only. This problem can be overcome by modifying suitably the coupling to the gauge fields at high energies. For example, adding the following operator
\begin{equation}
\dfrac{(\phi\overset{\text{\scriptsize$\leftrightarrow$}}{\mathcal{D}_\mu}\phi^\dag)(\phi^\dag\overset{\text{\scriptsize$\leftrightarrow$}}{\mathcal{D^\mu}}\phi) }{2\xi \phi^2+M_P^2}
\end{equation}
does the required job.

\section{Discussion and Conclusion}
\label{sec:DiscConcl}

To conclude, most of the efforts in solving the hierarchy problem up to date were associated with physics right above the EW scale (supersymmetry, dynamical Higgs or large extra dimensions). In this paper, we made an attempt to elaborate a different point of view, and to look at the Higgs mass as a low energy manifestation of non-perturbative phenomena associated with quantum gravity. We chose the Planck scale to start with, as it is the only scale appearing inevitably in any theory comprising the SM and GR. We then showed that, under specific assumptions about the high energy behaviour of the theory, it is possible to generate the observed EW scale via the instanton configuration that suppresses the value of $M_P$ by a necessary amount.

By construction, this mechanism does not require a fine-tuning among the coupling constants of the theory. In the model considered in sec. \ref{ssec:Poly}, the value of the ratio $\langle\varphi\rangle/M_P$ is mainly controlled by two parameters, $a_{HE}$ and $\delta$. Yet, already in this case, this value is degenerate in the parameter space, as fig. \ref{Fig:Degeneracy} demonstrates. We would like to note that, because $\delta$ appears in eq. (\ref{RhoBar}) with the small fractional power, the small values of it are required to satisfy eq. (\ref{CondOnDelta}). This fact is not related to the original hierarchy problem, and the smallness of $\delta$ does not bring about new interactions scales much below $M_P$. 

Speaking more generally, the non-perturbative mechanism considered here is by no means specific to the scalar-gravity model of sec. \ref{ssec:Poly}. For example, replacing the $(n=2)$-derivative operator (\ref{O_2_E}) with any $(n>2)$-operator or their linear combination yields qualitatively the same picture, since the essential features of the model (the finiteness of the singular shots and the finiteness of the instanton action) remain in game. Furthermore, it is possible to use other types of operators from the general Horndeski construction, that become total derivatives under variation with respect to the scalar field and, therefore, may regularize the divergence of the singular instanton. As for the polynomial operators used to enhance the scalar field source, it is clear that their form can also be different from that given in eqs. (\ref{F,G}), as long as the model (\ref{L0_J}) is restored at low field values.

On the other hand, it is clear that inclusion of higher-dimensional operators of the types different from those considered in sec. \ref{sec:Mechanism} can change the short-distance properties of the solution in the way incompatible with relation (\ref{CondOnW}). Here we would like to stress again that, instead of performing a barely possible analysis of euclidean classical configurations arising in effective theory of the Higgs field and gravity in full generality, we preferred to focus on particular examples on which we demonstrate the mere possibility of the existence of the desired non-perturbative effect. 

The singular instantons studied above turn out to be insensitive to the details of the model at low magnitudes of the scalar field. In particular, they are insensitive to the shape of the scalar field potential, at least for the reasonable values of the coupling constant $\lambda$. When applied to the real-world theory, this fact means that the non-perturbative effect coming with the inclusion of gravity does not depend on the physics much below the Planck scale. We cannot say definitely if this is a general feature of the mechanism, since we see no reason for the contribution to the instanton action from the low energy region to be negligible in general. Nonetheless, we find it intriguing that the observed hierarchy between the Fermi and the Planck scales could actually result purely from the features of quantum gravity \textit{above} the Planck scale.

Despite some arguments provided in favour of the exponential change of the scalar field variable (\ref{HiggsExp}), its implementation calls for further justification. Indeed, one would expect the correct answer for the vev of $\varphi$ to be independent of the choice of variables used at the intermediate steps of the calculation. One may note that the non-perturbative nature of the mechanism suggests that the effect caused by eq. (\ref{HiggsExp}) could be reproduced by certain resummation of perturbative contributions to the vev computed on top of the flat background; however, any detailed discussion of this question goes beyond the scope of the present paper.

A systematic treatment of fluctuations correcting the leading-order estimation (\ref{HiggsVEVSPA}) is not easy.\footnote{Some progress in a related problem of finding cosmological perturbations above a configuration with the singularity of the type (\ref{HEasympt}) was made in \cite{Gratton:1999ya}. } This task is important, however, as knowing the determinant in eq. (\ref{HiggsVEVSPA}) may clarify whether the exponential change of the field variable is indeed the preferred way to describe the dynamics at high energies. We leave to future work the investigation of different modes arising on top of the singular instanton, and their possible physical implications.

As was explained in sec. \ref{ssec:Summary}, the non-minimal coupling of the scalar field to gravity is one of the crucial parts of the mechanism. With this coupling preserved, one can expect instantons of a similar kind to exist in other scalar-tensor theories of gravity. In particular, being inspired by the idea of scale symmetry as a fundamental symmetry of Nature, it is tempting to study SI scalar-tensor theories that are reduced to the SM and GR in the process of spontaneous breaking of the scale invariance. A notable example of such theories is the Higgs-Dilaton model of \cite{GarciaBellido:2011de,Bezrukov:2012hx}. At the cost of an extra scalar degree of freedom --- the dilaton --- the model contains no scales at the classical level. The Planck mass is generated dynamically through the vev of the dilaton field. The Higgs-Dilaton model allows for reformulation of the hierarchy and the cosmological constant problems in terms of dimensionless parameters. It is interesting to see if it also allows for non-perturbative generation of the parameter determining the Higgs field vev, provided that the latter is classically zero. The investigation of this question is left for the subsequent publication \cite{Shaposhnikov:2018jag}.

\section*{Acknowledgements}

The authors thank Fedor Bezrukov and Javier Rubio for helpful discussions. This work was supported by the ERC-AdG-2015 grant 694896.  The work of M.S. and A.S. was supported partially by the Swiss National Science Foundation.

\bibliographystyle{elsarticle-num}
\bibliography{HI}

\end{document}